# Giant asymmetry of the longitudinal magnetoresistance in high-mobility two-dimensional electron gas on a cylindrical surface

A. B. Vorob'ev[1,2], K.-J. Friedland[1], H. Kostial[1], R. Hey[1], U. Jahn[1], E. Wiebicke[1],
Ju. S. Yukecheva[2], and V. Ya. Prinz[2]

[1]*Paul-Drude Institute for Solid State Electronics, Hausvogteiplatz 5-7, 10117 Berlin, Germany*

[2]*Institute of Semiconductor Physics, Acad. Lavrent'ev Ave. 13, 630090 Novosibirsk, Russia*

PACS numbers: 73.43.Fj, 73.43.Qt, 73.43.-f, 73.23.Ad

A giant asymmetry in the magnetoresistance was revealed in high-mobility, two-dimensional electron gas on a cylindrical surface. The longitudinal resistance along the magnetic-field gradient impressed by the surface curvature was found to vanish if measured along one of the edges of the curved Hall bar. If the external magnetic field is reversed, then the longitudinal resistance vanishes at the opposite edge of the Hall bar. This asymmetry is analyzed quantitatively in terms of the Landauer-Büttiker formalism.

## I. INTRODUCTION

Curved two-dimensional electron gases (2DEGs) hold much promise for investigation of new physical effects that appear both due to curvature itself[1] and due to modification of the external magnetic-field influence. The influence of a uniform external magnetic field on electrons in such gases is spatially non-uniform because only the normal-to-surface component of the magnetic field governs the Lorentz-force-related electron transport phenomena in 2DEGs. In terms of classical particle motion, the electron orbit radius undergoes changes as an electron moves along the curved surface, giving rise to quite complicated electron trajectories.[2] In quantizing magnetic fields, this non-uniformity, resulting in spatial curving of Landau levels (LL's),[3,4] manifests itself in magnetotransport characteristics which depend on the geometry of the experiment. Experimentally, free-standing shells with 2DEG shaped as open cylindrical surfaces of constant curvature can be fabricated using the strain-induced, three-dimensional micro- and nano-structuring method described in Ref. 5. In such objects the normal-to-surface component $B$ of the external magnetic field $B_0$ varies smoothly with the azimuth angle $\varphi$ in the plane normal to the generatrix of the cylindrical surface as $B=B_0\cos\varphi$ and changes it's sign. The gradient of the local magnetic field $B$ depends on the curvature radius $R$ of the surface as $(1/R)\partial B/\partial\varphi=-(B_0/R)\sin\varphi$ and amounts to $\sim 10^4$ T/cm for typical experimental conditions ($R\sim 10$ μm, $B_0\sim 10$ T). If the electric current is passed perpendicularly to the magnetic-field gradient, then spatially separated quasi-one-dimensional (1D) channels form along the cylindrical shell with 2DEG, which are expected to be spin-polarized due to the Zeeman splitting.[6] Magnetotransport of 2DEG in rolled-up microtubes for this current direction was studied experimentally by measuring magnetoresistance[7,8] and Hall effect.[9] Here we deal with the situation when the electric current flows along the periphery of the cylindrical shell where the 2DEG is located, i.e., parallel to the magnetic-field gradient. The classical diffusive transport in a conductive 2D strip was treated theoretically for magnetic field varying linearly along the strip.[10] The exact solution of Maxwell's equations yields a rather surprising result: the electric current flows along the strip near one of the strip edges depending on the sign of the magnetic-field gradient. This strongly nonuniform distribution of the current density results in an asymmetry of the longitudinal magnetoresistance in non-planar 2DEGs which was observed experimentally in 2DEGs on faceted surfaces[11] and in 2DEGs on cylindrical surface.[12] In the mentioned works, the longitudinal resistances for opposite orientations of the magnetic-field vector differed within a factor of 10.

In the present paper, we report on a giant asymmetry of the longitudinal magnetoresistance in high-mobility 2DEG on cylindrical surface under the conditions that the mean free path of the electrons compares with the dimensions of the sample. In this regime, close to the ballistic transport regime, in magnetic fields about 0.7-1.2 T the longitudinal resistance vanishes and in higher fields it stays very low if measured along one of the edges of the curved Hall bar, being of the order of the transversal resistance at the opposite edge. By reversing the direction of the external magnetic field the former low resistance edge becomes high resistance edge and vice versa. The ratio between the longitudinal resistances measured at one and the same edge of the Hall bar at opposite orientations of the magnetic field vector exceeds factor $10^3$.

## II. EXPERIMENT

A usual problem for free-standing thin (Al,Ga)As films results from the defects of the newly created surface which reduce the mobility considerably. We realized a high-mobility 2DEG in a free-standing thin film by using a GaAs quantum well with barriers consisting of AlAs/GaAs short-period superlattices (SPSL), in which heavy *X*-electrons exist. These heavy-mass *X*-electrons smooth the fluctuations of the scattering potential.[13] The influence of the defects of the newly created surface is reduced with this type of heterostructure. The multi-layered heterostructures were grown on a GaAs (001) substrate by molecular beam epitaxy (MBE). These structures comprised a 50-nm thick AlAs sacrificial layer and a strained multi-layered film with a nominal thickness of 192 nm to be released from the substrate. The cross-sectional view of the starting structure is shown in Figure 1a. The multi-layered film contains a 20-nm thick $In_{0.15}Ga_{0.85}As$ stressor and the stack of hereterostructure, including a central 13-nm-thick GaAs QW clad with δ-doped AlAs/GaAs SPSL-spacers. In the initial flat structure, the electron mobility along the high mobility [1$\bar{1}$0] direction and the electron sheet density are 118 $m^2$/(V·s) and $7.1 \times 10^{15}$ $m^{-2}$, respectively, resulting in a mean free path of electrons of about 16 μm.

The rolled-up Hall bars used in the present study were prepared by directional rolling of two-level-lithography defined mesastructures[14] as shown in Figs. 1b)-e). The photolithographic procedure comprised the following steps: (i) definition of a 6-terminal Hall bar and contact pads by shallow wet etching down to the lower δ-doped AlAs/GaAs SPSL; (ii) definition of the starting edge for rolling by deep etching down to the substrate; (iii) formation of Ohmic contacts on the flat contact pads; (iv) release of the patterned strained film by selective etching of the sacrificial layer, resulting in rolling of the film in the preset direction. Being detached from the substrate, the lithographically pre-patterned film directionally rolled along [100] direction in a scroll with visible outer curvature radius *R* of about 24 μm. After the rolling, each Hall bar represents a cylindrical sector attached to the large-area flat contact pads formed on the substrate. This sample was placed in a uniform magnetic field $B_0$ normal to the axis of the scroll. For such configuration, the local magnetic field *B* gradually varied along the Hall bar. The electric current in the Hall-bar also flows in the same direction and, hence, it was parallel to the magnetic-field gradient. The magnetotransport measurements were carried out in magnetic fields up to 14 T at 0.3 K using the conventional lock-in technique with an alternating current of 100 nA.

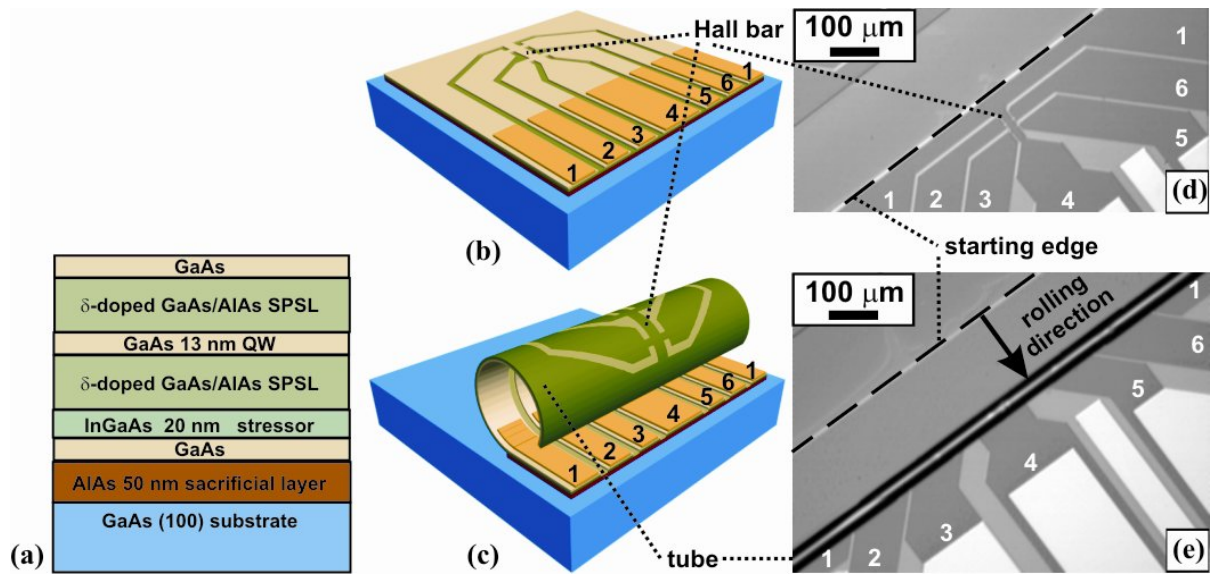

**Fig. 1.** (Color online) (a) Layer sequence of the MBE-grown multilayered heterostructure. (b-e) Arrangement of contacts to the scrolled Hall bar. At the left side: flat lithographically patterned Hall bar, at the right side: rolled-up Hall bar ((b) and (c) – sketches, (d) and (e) – plan-view photographs). Differences in the geometry of the leads and contact pads from ordinary Hall bars are distinctly seen: all the contact pads are positioned unilaterally with respect to the starting edge for rolling. The contact pads are indicated by numbers 1-6. Contacts 1 and 4 are current terminals, and contacts 2, 3, 5 and 6 are voltage probes.

### III. RESULTS

Figures 2a) and b) show the resistances $R_{xx}$ and $R_{xy}$ as functions of the magnetic field $B_0$ for different combinations of contacts in the rolled-up Hall-bar whose area in between the potential terminals was 16x16 μm. In the experiment illustrated by Fig. 2b the external magnetic field $B_0$ was normal to the surface roughly at the central point of the Hall bar, with the gradient of *B* changing its sign in the region between the potential contacts. In the latter case, the Hall effect in the free-standing film can be estimated as follows. The Hall voltage is determined by the local magnetic field $B=B_0\cos\varphi$ at the position of the transversal Hall terminals. With the separation $h=16$ μm between the transversal Hall terminals along the electric current path and with the rolling radius $R=24$ μm an angle of 38° for the circular arc can be estimated. For these terminals being positioned almost symmetrically about the direction of the external field $B_0$, the local field $B$ in the vicinity of either of the transversal Hall-terminals can be calculated as $B=B_0\cos(h/2R)\approx 0.95 B_0$. As a result, the electron density in the 2DEG determined from our Hall measurements turned out to be $7\times 10^{15}$ m$^{-2}$, this value being close to the sheet density of electrons in the initial planar structure. The electron mobility along the [100] direction in the rolled-up 2DEG was estimated to be near 20-30 m$^2$/(V·s), and the electron mean free path, near 3-4 μm. These values are significantly lower than the values measured along the [1$\bar{1}$0] direction in the flat sample, this being related to the mobility anisotropy due to the interface corrugations.[15] The rolling itself leaves the mobility in the given structures nearly unchanged. At the same time, a distinct negative bend resistance was observed in cross junctions formed by the opposite terminals of the rolled-up Hall bar with area 8x8 μm (made from the same MBE-grown structure) in between the potential terminals. This indicates ballistic transport of carriers at that distance and a longer mean free path than that extracted from the Hall measurements of the 2DEG with highly anisotropic mobility (a detailed description and an analysis of the latter experiment will be published elsewhere[16]). It is worth to note that the curves of $R_{xx}(B)$ in Fig. 2b (for either contact pair and for both magnetic-field directions) show clear SdH oscillations which start already from relatively weak fields together with a vanishing longitudinal resistance at the minima of the oscillations in accord with features due to the spin-split integer QH states. The curves of $R_{xy}(B)$ display well-pronounced Hall plateaus corresponding to quantized values of $h/ie^2$. These features are characteristic for high-mobility 2DEG.

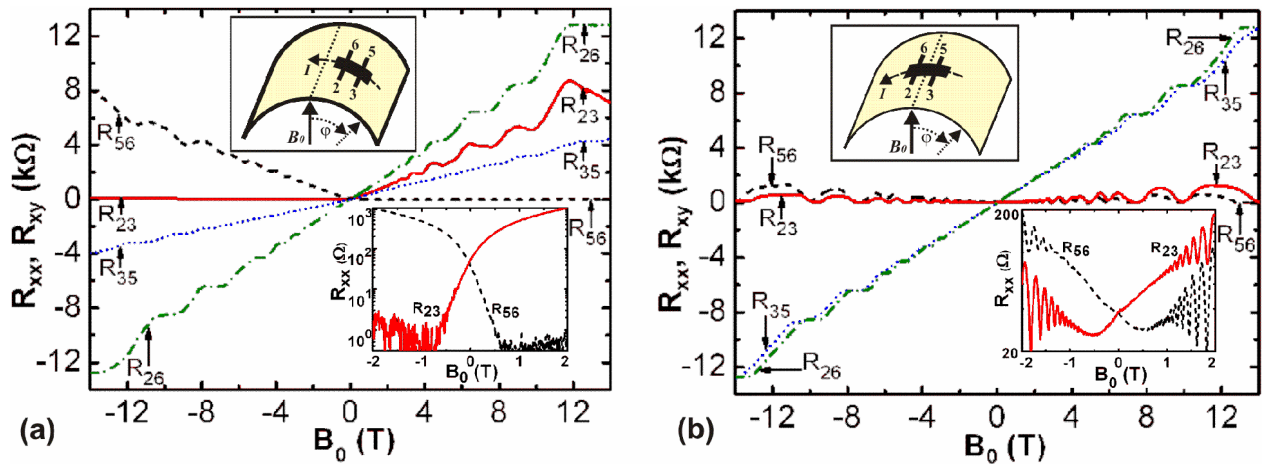

**Fig. 2.** (Color online) Longitudinal and transverse resistances of the cylindrical 2DEG vs magnetic field measured for different orientations of one and the same Hall bar with respect to the magnetic field direction. The external magnetic field is oriented normally to the surface (a) at a position somewhere outside the region between the contacts, (b) - roughly at the central point of the Hall bar. The upper insets show the experiment geometries, and the lower insets show longitudinal resistances vs magnetic field on a logarithmic scale. The direction of the current I and the direction of the external magnetic field $B_0$ are shown with the dashed and solid arrow, respectively. The resistances measured between different pairs of probes 2, 3, 5 and 6 are indicated with subscripts.

Quite a different magnetic-field dependence of the longitudinal resistance arises in that case for which the external magnetic field $B_0$ is oriented normally to the surface at a position somewhere outside the region between the contacts (see the upper inset in Fig. 2a). In this case the local field $B$ varies much stronger yet monotonically in the region between the potential contacts. Here, the longitudinal resistance $R_{xx}$ in the 2DEG on cylindrical surface becomes strongly asymmetric with respect to the orientation of the external field $B_0$. Interesting and surprising features of this asymmetry in the longitudinal magnetoresistance are: (1) rapid vanishing of the longitudinal resistance with the magnetic-field strength; (2) giant ratio between the longitudinal resistances measured at both pairs of contacts 2-3 and 5-6 for opposite directions of the magnetic field (this ratio is greater than 1000 in a magnetic field of 1 T); (3) high sensitivity of the asymmetry even to slightest misalignments of the rolled-up sample from the symmetric position with respect to the magnetic-field vector.

## IV. DISCUSSION

In order to explain this enormous asymmetry of the longitudinal magnetoresistance we note, that for these high mobility structures Landau quantization takes place already at relatively low magnetic fields (below 1T). In planar 2DEG's the electric current in the QH effect mode is carried by edge channels resulting from the LL's that intersect the Fermi level. For 2DEG's on cylindrical surface, when the magnetic field is non-uniform along the Hall bar, bending of LL's gives rise to 1D channels propagating normally to the direction of the magnetic-field gradient, which form in those places where the Fermi level crosses the LL's (Fig. 3a). In the interior of the sample these channels merge together with the edge channels, forming closed contours[4] along which the electrons that leave the injecting contact return to it. Apparently, the location of the channels depends on the strength of the magnetic field. As the strength of the magnetic field increases, the channels move closer to the contact where the magnetic-field intensity is low. In the present model, in a sufficiently strong magnetic field, when the Fermi level is below the lowest LL over the most to the part of the sample, all electrons leaving the contact would return to it, which would result in complete blockage of the electric current across the sample.

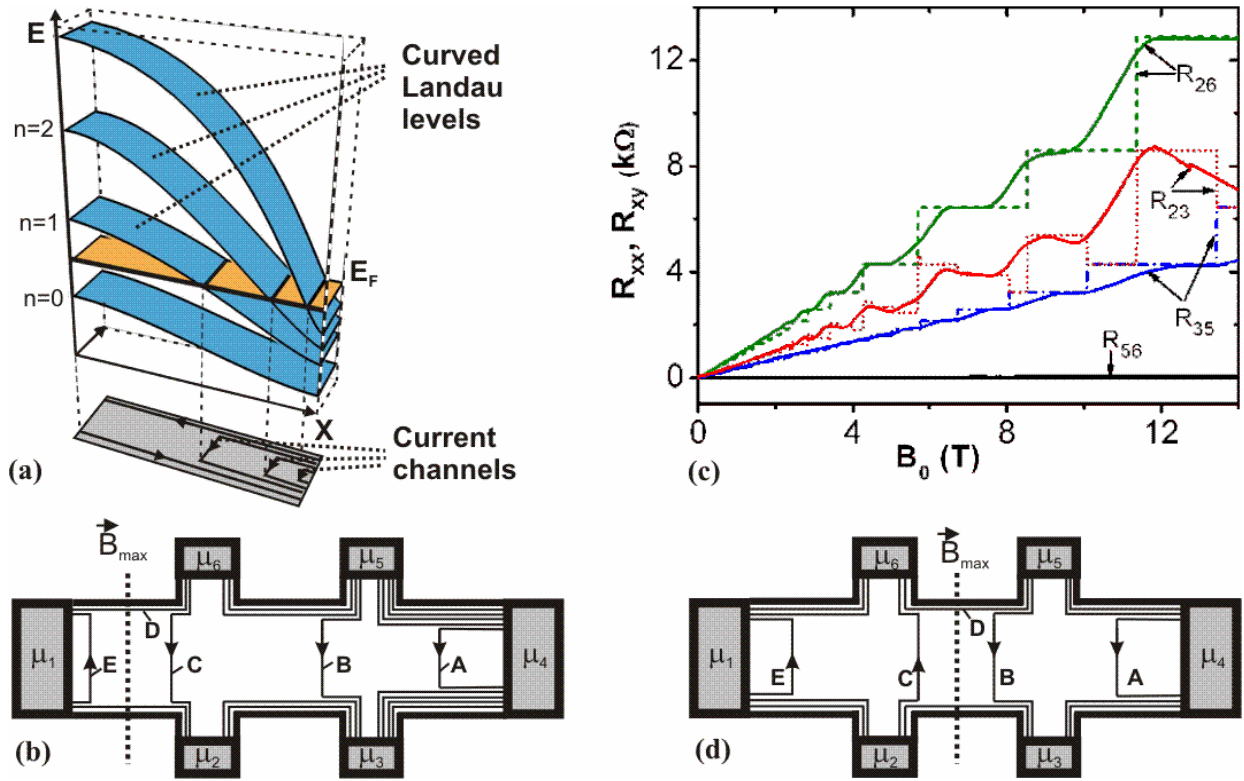

**Fig. 3.** (Color online) (a) Curved LL's in a magnetic field smoothly varying along the Hall bar. (b) Schematic representation of the 1D channels in the Hall bar for the case in which the normal magnetic-field component $B$ monotonically varies in the region between the potential contacts (see the upper inset in Fig. 2a). (c) Experimental (solid lines) and calculated (dashed, dotted and dashed-dotted lines) resistances $R_{26}$, $R_{23}$, $R_{35}$, $R_{56}$. (d) Schematic representation of the 1D channels in the Hall bar for the case in which the gradient of $B$ changes its sign in the region between the potential contacts (see the upper inset in Fig. 2b).

Following Ref. 4, all of the 1D channels can be subdivided into several types (Fig. 3b and 3d). A and E are the channels that return into the current contact not reaching the nearest pair of Hall contacts; B are the channels that enter the nearest pair of the Hall contacts, C are the channels that enter one pair or both pairs of Hall contacts depending on the position where the external magnetic field $B_0$ is normal to the surface, and D are the channels that enter both pair of potential contacts and reach the second current contact. The contact patterns in Fig. 3b and 3d are simplified, i.e. the long side arms between the curved Hall bar and the flat metallic pads in the real sample (see Fig. 1c) are not shown. The simple contact patterns are applicable because, throughout the whole range of external magnetic fields at the given density of charge carriers in the 2DEG, at least one edge channel passes from the Hall bar to the metallic contact pads. This channel brings to the metal pads an already equilibrium electrochemical potential due to scattering events between channels of different types that happen over a long distance (several times 100 microns), and due to the passage of the chan-nels under consideration through the region with zero local field. Indeed, magnetic barriers are the same for side arms 2 and 6 as well as for side arms 3 and 5. Nonetheless, the potential levelling between channels of different types within each side arm allows us to observe Hall resistances corresponding to the local values of magnetic field $B$ and the asymmetry of longitudinal magnetoresistance. The redistribution of the edge channels occupancy due to the finite equilibration length will be discussed in detail elsewhere.[17]

Let us consider the case, in which the magnetic field is perpendicular to the layer at a position outside the region between the contacts with chemical potentials $\mu_2$, $\mu_3$, $\mu_5$, $\mu_6$ and the electrons flow out of the left current electrode with a chemical potential $\mu_4$ (Fig. 3b). All of the channels coming into the probe with the potential $\mu_5$ leave the current electrode with the potential $\mu_4$, then, $\mu_4=\mu_5$. Likewise, $\mu_5=\mu_6$ and, hence, the longitudinal resistance measured on the upper face of the sample is $R_{56}=(\mu_6-\mu_5)/eI=0$. The vanishing of $R_{xx}$ on one side of the bent Hall bar is of the same nature as the vanishing of $R_{xx}$ in planar samples under QH-effect conditions. Yet, in contrast to the planar case, here the longitudinal resistance turns to zero throughout the whole interval of quantizing magnetic fields. However, on the opposite edge of the sample, the potentials $\mu_2$ and $\mu_3$ at the voltage probes generally differ from each other; the difference $(\mu_3-\mu_2)$ is determined by addition of the channels B to the channels D and C; as a result, the longitudinal resistance, $R_{23}=(\mu_3-\mu_2)/eI$, is non-zero in the whole interval of quantizing magnetic fields. In the experiment, we observed vanishing of the longitudinal resistance $R_{56}$ at magnetic fields 0.7 to 1.2 T. At higher fields the magnetoresistance asymmetry remains high $(R_{xx}(B)/R_{xx}(-B))>100)$, in spite of small deviation of $R_{56}$ from zero with increasing external magnetic field: the resistance $R_{56}$ grows in value starting from the noise level (0.2-0.3 Ohm) at ~1 T up to 20-25 Ohm at ~ 14 T, displaying strongly suppressed SdH oscillations. Possible cause for this small deviation at high fields is some imperfection of the contacts and insufficiently high input impedance of standard phase-sensitive voltmeters.[18] In the experiment presented in Fig. 2a at high magnetic field (for "positive" orientation of $B_0$) some inconsistency is observed between measured values of resistances and the Kirchhoff's law $R_{26}= R_{23}+ R_{35}+ R_{56}$. It is caused by mechanical displacement of the free-standing part of the sample, which sometimes happens in strong magnetic fields.

The longitudinal and Hall resistances were calculated using the Landauer-Büttiker approach[19] which is well established for the 1D conductance in the QH regime. Taking into account that the current $I=e/h\Delta\mu$ flows in each of the current channels resulting from spin-split LL's, we sum up the currents incoming and outgoing from each of the contacts. The Landauer-Büttiker equations for the situation of interest can be represented in matrix form (see, e.g., Ref. 20) as follow:

$$\begin{pmatrix} I \\ 0 \\ 0 \\ -I \\ 0 \\ 0 \end{pmatrix} = \frac{e}{h} \begin{pmatrix} -D & 0 & 0 & 0 & 0 & D \\ D & -C-D & 0 & 0 & 0 & C \\ 0 & C+D & -B-C-D & 0 & B & 0 \\ 0 & 0 & B+C+D & -B-C-D & 0 & 0 \\ 0 & 0 & 0 & B+C+D & -B-C-D & 0 \\ 0 & 0 & 0 & 0 & C+D & -C-D \end{pmatrix} \cdot \begin{pmatrix} \mu_1 \\ \mu_2 \\ \mu_3 \\ \mu_4 \\ \mu_5 \\ \mu_6 \end{pmatrix}$$

From here, the longitudinal and Hall resistances can be calculated as

$$R_{23} = \frac{\mu_3 - \mu_2}{eI} = \frac{h}{e^2} \cdot \frac{B}{(B+C+D)\cdot(C+D)}, \quad R_{56} = \frac{\mu_6 - \mu_5}{eI} = 0 \,, \quad (1)$$

$$R_{26} = \frac{\mu_6 - \mu_2}{eI} = \frac{h}{e^2} \cdot \frac{1}{C+D}, \quad R_{35} = \frac{\mu_5 - \mu_3}{eI} = \frac{h}{e^2} \cdot \frac{1}{B+C+D} \,. \quad (2)$$

The number of the current channels (or the number of locally occupied spin-split LL's) can be determined using the electron concentration $n_S$ and the orientation of the sample relative to the external magnetic field:

$$B + C + D = Int\left[\frac{n_s \cdot h}{eB_1}\right], \quad C + D = Int\left[\frac{n_s \cdot h}{eB_2}\right] \quad (3)$$

where $Int[]$ signifies taking the integer part, $B_1$ and $B_2$ are the local magnetic fields near the Hall-contact pairs 3-5 and 2-6, respectively. These fields can be found from the values of the external magnetic field at one and the same Hall voltage between different pairs of Hall contacts (Fig. 2a), taking into account the known tube radius of 24 μm and the separation between the two contact pairs of 16 μm. We calculated the longitudinal and Hall resistances versus the external magnetic field using equations (1)-(3) considering resolved spin splitting of LL's at external fields above 7 T. At lower magnetic fields the LL's were considered as spin-degenerated levels. The calculated curves match well with experimental data (Fig. 3c). The difference in the shapes of the curves results from possible inaccuracy in sample geometry definition and neglecting the broadening of LL's $\Gamma = \hbar\sqrt{\dfrac{2\varpi_c}{\pi\tau}}$ (here, $\omega_c = eB/m^*$ is the cyclotron frequency and $\tau$ is the scattering time). The LL broadening increases the scattering probability of electrons between neighboring 1D channels. Each scattering event can cause a transition of an electron into a neighboring channel propagating along another path (from the channel A into the channel B in Fig.3b, for instance); this situation fundamentally differs from the planar 2DEG in which the channels located near one side of the sample all propagate in the same direction. This means that in the curved 2DEG the interchannel scattering plays a more important role for the electron transport as compared to the planar 2DEG.

The current-path pattern for the case in which the gradient of $B$ changes its sign in the region between the potential contacts is shown in Fig. 3d. Now, the channels extending in opposite directions from either of the current terminals contribute to the longitudinal resistances. This situation was analyzed in Ref. 21, where the following expressions for the resistances were derived:

$$R_{23} = \frac{\mu_3 - \mu_2}{eI} = \frac{h}{e^2}\cdot\frac{B}{(B+D)\cdot D}; \quad R_{56} = \frac{\mu_6 - \mu_5}{eI} = \frac{h}{e^2}\cdot\frac{C}{(C+D)\cdot D};$$

$$R_{26} = \frac{\mu_6 - \mu_2}{eI} = \frac{h}{e^2}\cdot\frac{1}{C+D}; \quad R_{35} = \frac{\mu_5 - \mu_3}{eI} = \frac{h}{e^2}\cdot\frac{1}{B+D}.$$

It is clear that, depending on the position where the external magnetic field $B_0$ is normal to the surface, an asymmetry in the longitudinal magnetoresistance can be observed; yet, here, in contrast to the former case, the resistance does not vanish in the whole interval of magnetic fields.

### V. CONCLUSION

To summarize, structures with a high-mobility 2DEG in rolled-up semiconductor films were fabricated to examine the magnetotransport properties of 2DEG on cylindrical surface. A strong magnetoresistance asymmetry in such structures was observed for the direction of the electric current parallel to the magnetic-field gradient (the longitudinal resistance vanishes if measured along one of the edges of the curved Hall bar). This asymmetry was explained using the Landauer-Büttiker approach.

### ACKNOWLEDGEMENTS

One of the authors (A.V.) greatly appreciates the support from INTAS (Grant No. 04-83-2575) and from the Russian Science Support Foundation.